\documentclass[a4paper,12pt]{article}

\usepackage{physics}
\usepackage{graphicx}
\usepackage{float}
\usepackage{caption}
\usepackage{array}
\usepackage{algorithm,algpseudocode}
\usepackage{forest}
\usepackage{qcircuit}
\usepackage{wasysym}
\usepackage{amssymb}
\usepackage{multirow}
\usepackage{xparse}
\usepackage{siunitx}
\usepackage{authblk}
\author[1,2]{Doha A. Rizk}
\author[1,2,3]{Ahmed Younes}
\affil[1]{Department of Mathematics and Computer Science, Faculty of Science, Alexandria University, Egypt.}
\affil[2]{Academy of Scientific Research and Technology, Cairo, Egypt.}
\affil[3]{School of Computer Science, University of Birmingham, Birmingham, UK.}
\makeatletter
\NewDocumentCommand{\LeftComment}{s m}{%
	\Statex \IfBooleanF{#1}{\hspace*{\ALG@thistlm}}\(\triangleright\) #2}
\makeatother

\makeatletter
\newenvironment{breakablealgorithm}
{
	\begin{center}
		\refstepcounter{algorithm}
		\hrule height.8pt depth0pt \kern2pt
		\renewcommand{\caption}[2][\relax]{
			{\raggedright\textbf{\ALG@name~\thealgorithm} ##2\par}%
			\ifx\relax##1\relax 
			\addcontentsline{loa}{algorithm}{\protect\numberline{\thealgorithm}##2}%
			\else 
			\addcontentsline{loa}{algorithm}{\protect\numberline{\thealgorithm}##1}%
			\fi
			\kern2pt\hrule\kern2pt
		}
	}{
		\kern2pt\hrule\relax
	\end{center}
}
\makeatother
\begin{document}
	\title{Quantum Algorithm for Quantum State Discrimination via Partial Negation and Weak Measurement}
	\date{\today}
	\maketitle
	\begin{abstract}
		The quantum state discrimination problem is to distinguish between non-orthogonal quantum states. This problem has many applications in quantum information theory, quantum communication and quantum cryptography. In this paper a quantum algorithm using weak measurement and partial negation will be proposed to solve the quantum state discrimination problem using a single copy of an unknown qubit. The usage of weak measurement makes it possible to reconstruct the qubit after measurement since the superposition will not be destroyed due to measurement. The proposed algorithm will be able to determine, with high probability of success, the state of the unknown qubit and whether it is encoded in the Hadamard or the computational basis by counting the outcome of the successive measurements on an auxiliary qubit.
	\end{abstract}
     \section{Introduction}
     Quantum state discrimination is the process of discriminating among a known set of possible states $\ket{\psi_i}$,\textit{i}=0,1,2,3,...,\textit{n}, each has been prepared with a known probability $p_i$. In quantum mechanics, a collection of orthogonal states can be perfectly discriminated, while one have to accept an error or inconclusive results if the quantum states are not orthogonal, for example there is no process that can distinguish between the states $\ket{0}$ and $\frac{1}{\sqrt{2}}$($\ket{0}$+$\ket{1})$ with certainty. The problem of discriminating between non-orthogonal states is fundamental to the quantum theory of communication and quantum cryptography\cite{Bennett1992,Enk2002}. An eavesdropper will not be able to both apply measurement and send the message to the intended receiver without the recognition of his presence as shown in the BB84 protocol\cite{Bennett2014} where the role of the No-cloning theorem comes to be crucial \cite{Wootters1982}.
     
     The quantum state discrimination problem can be demonstrated as two parties Alice and Bob communicating through a quantum channel, where Alice selects a state randomly from a given ensemble of quantum states $\ket{\psi_i}$  with a priori probabilities $p_i$; $0 \leqslant i \leqslant n$; which is known to both of them, the state is then encoded and sent to Bob. Bob is then required to guess the state through a suitable quantum measurement to get the intended message \cite{Stephen2009}. This is considered as a non-trivial problem, since for an ensemble of non-orthogonal quantum states, quantum measurements are not able to give full information about the prepared system \cite{Bae2015}.  
     
     Optimization schemes have been presented as an alternative option to discriminate non-orthogonal quantum states. Minimum error state discrimination strategy introduced by Helstorm \cite{Helstrom1976}, where the observer gives conclusive results with minimum average error. Unambiguous state discrimination strategy was first suggested by Ivanovic \cite{Ivanovic1987}, where it allows inconclusive results, since the observer cannot  conclude the received state correctly with $100\%$ success. Later Ivanovic \cite{Ivanovic1987}, Dieks \cite{Dieks1988} and Peres \cite{Peres1988} optimized the discrimination strategy by minimizing the probability of occurance of inconclusive results, and they obtained a minimum value called Ivanovic-Dieks-Pers (IDP) limit such that, their priori probabilities are uniform. Jager and Shimony \cite{Jaeger1995} then extended the solution of the unambiguous discrimination to an arbitrary number of prior probabilities. Unambiguous discrimination between linearly independent quantum states approach have been introduced by Chefels \cite{Chefles1998} where it was shown that measurement of non-orthogonal states gives non-zero probability of inconclusive results and that only independent quantum states could be unambiguously discriminated. Later Chefels \cite{Chefles2001} introduced unambiguous discrimination between linearly dependent states, were multiple copies of a state were used to make the unambiguous discrimination of linearly dependent states possible, however using multiple copies of a qubit results in a high cost. Ballester et al. \cite{Ballester2008} presented a state discrimination problem using post-measurement information, where quantum storage was required. Fanizza et al. \cite{Fanizza2019}, suggested an optimal universal learning quantum machine that aimed to solve the discrimination problem of a $d$-level quantum system which is known to be in one of $d$-states with arbitrary a priori probabilities. The quantum machine is given a partial information about the states, and the machine is trained by preparing $n$ qudits for each state. Sequential state discrimination strategy has been proposed by Bergou et al. \cite{Bergou2013} where multiple observers take place, and the aim is to get the maximum success probability. Later Zhang et al. \cite{zhang2018} introduced a solution to this problem for two pure qubits with arbitrary a priori probabilities. Namkung et al. \cite{Namkung2018} then, extended the sequential state discrimination approach to $N$ linearly independent pure quantum state. Peng et al. \cite{peng2020} presented an optimal weak measurement method to maximize the mutual information among multi-party quantum communication, an experimental work have been carried out by a cascaded measurement interferometer that performed weak measurements, where the total accessible information gained was found to be approximately equal 1. Xin L$\ddot{u}$ \cite{Lu2020} determined the upper bound success probability of the minim-error state discrimination defined in \cite{Helstrom1976} using the wave particle duality. The idea of incompatibility of quantum measurement gives an advantage in the quantum state discrimination, where \cite{Carmeli2018,Carmeli2019,Skrzypczyk2019} showed that quantum incompatibility increases the guessing probability of the prior measurement information to the post measurement information, where Alice gives partial information to Bob which can either be sent before Bob has performed the measurement or after he arranges the measurement.
     
     The aim of this paper is to propose a quantum algorithm using weak measurement and partial negation in order to reconstruct a qubit after measurement, and hence distinguish between different quantum states. A measurement-based quantum random walk approach proposed in \cite{Younes2017} is used for reading a single unknown qubit $\ket{\psi}$ in order to minimize the disturbance in the superposition, where it will be shown that it is possible to determine whether the measurement process is in the computational or the Hadamard bases with an approximate percentage 75\% accuracy using a single qubit.
     
     The paper is structured as follows, Section 2 defines the problem to be solved by the proposed algorithm. Section 3 illustrates the partial negation operator that is used to create a weak entanglement between an auxiliary qubit and the unknown qubit. Section 4 presents the measurement-based quantum random walk approach used in the proposed algorithm. Section 5 proposes an algorithm to distinguish between different quantum states. Section 6 discusses the obtained results. Finally, Section 7 concludes the paper.
     
     \section{Problem Statement}
     Given a qubit $\ket{\psi}$ which is promised to be in one of the following four states $\ket{0}$, $\ket{1}$, $\ket{+}$ and $\ket{-}$ where, $\ket{\pm}=\frac{1}{\sqrt{2}}(\ket{0}\pm\ket{1})$. It is required to determine the state of $\ket{\psi}$ using a single copy and whether it is encoded in the Hadamard or the computational basis.
     The probabilities that the unknown qubit $\ket{\psi}$ is in one of the previously mentioned four states when measured in the computational or the Hadamard basis is shown in Table (1).  
     
     \begin{center}
     	\begin{tabular}{ |c|c|c|c| } 
     		\hline
     			$\ket{\psi}$ & \begin{tabular}[c]{@{}c@{}}Measurement using\\ computational\\ basis\end{tabular} & \begin{tabular}[c]{@{}c@{}}Measurement using\\ Hadamard\\ basis\end{tabular}\\ \hline
     			$\ket{0}$ & $100\%$ & $50\%$ \\ \hline 
     			$\ket{1}$ & $100\%$ & $50\%$ \\ \hline 
     			$\ket{+}$ & $50\%$ & $100\%$ \\ \hline 
     			$\ket{-}$ & $50\%$ & $100\%$ \\ \hline
     	\end{tabular}
     	\captionof{table}{Probailities that an unknown qubit $\ket{\psi}$ is in state $\ket{0}$, $\ket{1}$, $\ket{+}$ or $\ket{-}$ when measured in the computational or the Hadamard basis.}
     	
     \end{center}
     
     Given an arbitrary qubit $\ket{\psi}=\alpha\ket{0}+\beta\ket{1}$ it can be expressed in the Hadamard basis ($\ket{+}, \ket{-}$) in the following way,
     \begin{equation}
     \ket{\psi}=\alpha\ket{0}+\beta\ket{1}=\frac{\alpha+\beta}{\sqrt{2}}\ket{+}+\frac{\alpha-\beta}{\sqrt{2}}\ket{-}.
     \end{equation}
      This problem has a special importance in Quantum key distribution (QKD). Many QKD protocols have been introduced such as  BB84 \cite{Wootters1982}, E91 \cite{ekert1991quantum}  and B92 \cite{bennett1992experimental} protocols, where the security of the system allows an Eavesdropper to determine the classical data with $50\%$ correctness \cite{gisin2002quantum}.
     
     \section{Partial Negation Operator}
     In quantum computing, a gate can be represented by a unitary matrix $U$, where some gates act on a single qubit, others act on $n$ qubits, such that $n>1$. One of the gates that act on a single qubit is the negation $\sigma_x$ gate, which is equivalent to the classical NOT gate, i.e. it inverts $\ket{0}$ to $\ket{1}$ and $\ket{1}$ to $\ket{0}$.The unitary matrix of the $\sigma_x$ gate can be represented as,
     \begin{equation}
     \sigma_x = \begin{bmatrix} 0 & 1 \\ 1 & 0\end{bmatrix}.
     \end{equation}
     The rotation gate $R_x(\theta)$ can be defined using the following equation \cite{soeken2013},
     \begin{equation}
     R_x(\theta) = e^{-i \frac{\theta}{2} \sigma_x} = cos(\frac{\theta}{2})I- i\; sin(\frac{\theta}{2})\sigma_x,
     \end{equation}
     such that the angle $\theta \in \mathbb{R}$ and $I$ is an identity operator.
     
     The $\sigma_x$ gate, is constructed with a rotation angle $\theta=\pi$ as shown below,
     \begin{equation}
     \sigma_x = e^{\frac{i\pi}{2}}R_x(\pi),
     \end{equation}
     where, $e^{\frac{i\pi}{2}}$ is a global phase shift \cite{soeken2013}.
     
     The $t^{th}$ root of the $\sigma_x$ gate is defined using an operator $V$ as follows,
     \begin{equation}
     V=\sqrt[t]{\sigma_x} = e^{\frac{i\pi}{2t}}R_x(\frac{\pi}{t}),  
     \end{equation}
     such that, $V$ can be represented as a unitary matrix in the following form,
     \begin{equation}
     V = \sqrt[t]{\sigma_x} =\frac{1}{2}\begin{bmatrix} 1+k & 1-k \\ 1-k & 1+k\end{bmatrix},
     \end{equation}
     where $k=\sqrt[t]{-1}$.
     
     A controlled $V$-gate is used to define an operator $P$, such that the $V$-gate is applied conditionally for $d$ times on an auxiliary qubit denoted as $\ket{ax}$ and initialized to the state $\ket{0}$. The number of times the $V$-gate is applied on $\ket{ax}$ depends on the 1-density of a vector $\ket{b_1b_2b_3....b_n}$ which is a set of qubits $\ket{b_1}$, $\ket{b_2}$,..,$\ket{b_n}$, and the 1-density denoted as $d$ is the number of qubits in the state $\ket{1}$ i.e. $b_i=1$ in the vector $\ket{b_1b_2b_3....b_n}$ and $d\leqslant n$, as shown in figure (1). If $d=n$ then, $\sigma_x$-gate will be applied on $\ket{ax}$ \cite{Younes2015}.
     
     Applying an operator $V$ for $d$ times can be defined using an operator $V^d$ as follows,
     \begin{equation}
     V^d = \frac{1}{2}\begin{bmatrix} 1+k^d & 1-k^d \\ 1-k^d & 1+k^d\end{bmatrix}.
     \end{equation} 
     \begin{center}
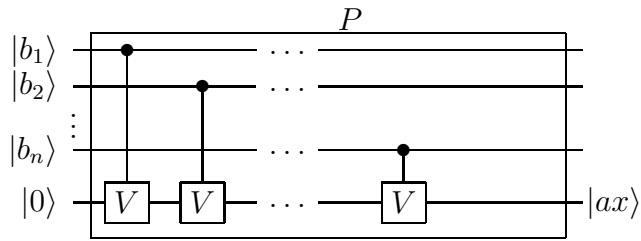

     	\begin{tabular}{c}
     		\Qcircuit @C=1em @R=.9em {
     			& & & & & & 	& \mbox{$P$}\\
     			& \lstick{\ket{b_1}} & \ctrl{4} &\qw &\qw & \dots & & \qw &\qw &\qw &\qw &\qw &\qw &\qw\\
     			& \lstick{\ket{b_2}}  & \qw & \ctrl{3} & \qw & \dots & &\qw &\qw &\qw &\qw &\qw &\qw &\qw\\
     			&\vdots\\
     			& \lstick{\ket{b_n}} & \qw & \qw & \qw & \dots & & \qw & \ctrl{1} & \qw &\qw  &\qw &\qw &\qw\\
     			& \lstick{\ket{0}} & \gate{V} & \gate{V} & \qw & \dots & & \qw & \gate{V}&\qw &\qw &\qw &\qw&\qw &\ket{ax}\gategroup{2}{3}{6}{13}{.9em}{-}
     		} 
     	\end{tabular}
     	\captionof{figure}{Quantum circuit for the $P$ operator representing $n$ number of controlled $V$-gates being applied conditionally on $\ket{ax}$.}
     \end{center}
     
     The $P$ operator can be defined in equation (8) as follows,
     \begin{equation}
     P = ControlV(b_1,ax)ControlV(b_2,ax).....ControlV(b_n,ax),
     \end{equation}
     where, $ControlV(b_i,ax)$ is a representation for the controlled $V$-gate such that the control is a qubit  $\ket{b_i}$ and the target is the auxiliary qubit $\ket{ax}$.  
     Applying $P$ on $n+1$ qubits register, i.e. $\ket{b_1b_2b_3....b_n}$ and an auxiliary qubit $\ket{ax}$, then
     \begin{equation}
     P(\ket{b_1b_2b_3....b_n}\otimes\ket{0}) = \ket{b_1b_2b_3....b_n}\otimes(\frac{1+k^d}{2}\ket{0} + \frac{1-k^d}{2}\ket{1}),
     \end{equation}  	
     and the probabilities of finding $\ket{ax}$ in state $\ket{0}$ or $\ket{1}$ when measured is given by,
     \begin{equation}
     Pr(\ket{ax} = \ket{0})= \abs{\frac{1+k^d}{2}}^2=\cos^2{(\frac{d\pi}{2t})},
     \end{equation}
     \begin{equation}
     Pr(\ket{ax} = \ket{1})= \abs{\frac{1-k^d}{2}}^2=\sin^2{(\frac{d\pi}{2t})}.
     \end{equation}
     \section{Measurement Based Quantum Random Walk Approach}
     Measurement based quantum random walk is an algorithm proposed in \cite{Younes2017} to read the information content of an unknown qubit $\ket{\psi}$, using weak measurement in order to minimize the disturbance introduced to the superposition.
     
     Quantum measurement, or sharp measurement is a projective measurement that causes the superposition to collapse to one of its possible states in a probabilistic manner, thus destroying the superposition which can’t be reconstructed again, because projective measurement is an irreversible operation. Alternatively, weak or unsharp measurement can be conceived, which is achieved when the coupling of the apparatus with the system is weak \cite{Das2014}. Accordingly, only limited disturbance can be introduced to the superposition, and hence reconstructing the qubit after measurement.     
     
     Weak measurement can be applied on an unknown qubit $\ket{\psi}$, by preparing the entanglement between the unknown qubit and an auxiliary qubit $\ket{ax}$. A register $(\ket{\psi}\otimes\ket{ax})$ is prepared by appending $\ket{ax}$ to $\ket{\psi}$ as shown in figure (2), where $\ket{\psi}=\cos(\phi)\ket{0}+\sin(\phi)\ket{1}$ and $\ket{ax}$ is initialized to the state $\ket{0}$. A sharp measurement is then applied on $\ket{ax}$, causing the effect of weak measurement on the unknown qubit $\ket{\psi}$ entangled with it.
     \begin{center}
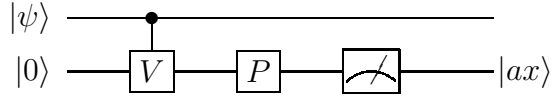

     	\begin{tabular}{c}
     		\Qcircuit @C=1em @R=.9em {
     			\lstick{\ket{\psi}} &\qw &\ctrl{1} &\qw &\qw  &\qw &\qw &\qw &\qw&\qw\\
     			\lstick{\ket{0}} &\qw &\gate{V} &\qw &\gate{P} &\qw&\meter &\qw&\qw&\qw &\ket{ax}
     		}
     	\end{tabular}
     	\captionof{figure}{Quantum circuit for an entanglement between the two qubits $\ket{\psi}$ and  $\ket{ax}$, and applying a controlled $P$ operator on the system.}
     \end{center}
 	
 	The strength of the weak measurement can be controlled by appending an arbitrary number of dummy qubits $\mu$ to the system, such that all $\mu$ qubits are initialized to the state $\ket{1}$ as shown in figure (3). The amount of disturbance the weak measurement causes to the superposition of the unknown qubit, depends on the number of the dummy qubits used, i.e. as the number of dummy qubits decrease, the strength of the weak measurement will be maximized, which will introduce more disturbance to the superposition and vice versa.  
 	
     \begin{center}
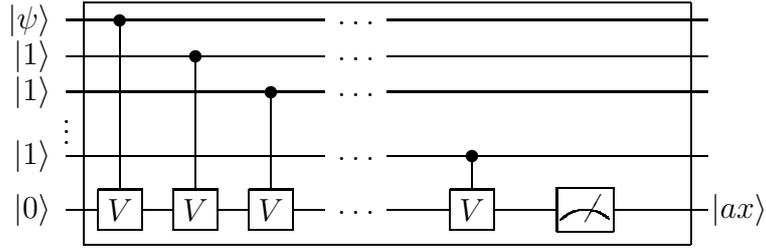

     	\begin{tabular}{c}
     		\Qcircuit @C=1em @R=.9em {
     			& & & & & & 	\\
     			& \lstick{\ket{\psi}} & \ctrl{5} &\qw &\qw &\qw & \dots & & \qw &\qw &\qw &\qw &\qw &\qw &\qw\\
     			& \lstick{\ket{1}} & \qw & \ctrl{4} &\qw &\qw & \dots & & \qw &\qw &\qw &\qw &\qw &\qw &\qw\\
     			& \lstick{\ket{1}}  & \qw & \qw & \ctrl{3} & \qw & \dots & &\qw &\qw &\qw &\qw &\qw &\qw &\qw\\
     			&\vdots\\
     			& \lstick{\ket{1}} & \qw & \qw & \qw & \qw & \dots & & \qw & \ctrl{1} & \qw &\qw  &\qw &\qw &\qw\\
     			& \lstick{\ket{0}} & \gate{V} & \gate{V} & \gate{V} & \qw & \dots & & \qw & \gate{V}&\qw &\meter&\qw&\qw&\qw &\ket{ax}\gategroup{2}{3}{7}{14}{.9em}{-}
     		} 
     	\end{tabular}
     	\captionof{figure}{Quantum circuit for the entanglement between $\ket{\psi}$ and $\ket{ax}$, with a $P$ operator applied on $\ket{ax}$ and $\mu$ qubits initialized to the state $\ket{1}$.}
     \end{center}
     
     Appending $\mu$ number of dummy qubits to the system, the register will be as shown,
     \begin{equation}
     \begin{aligned}
     P(\ket{\psi}\otimes\ket{0})&= P(\ket{\psi}\otimes\ket{1}^{\otimes\mu}\otimes\ket{0})\\
     &=P((\cos(\phi)\ket{0}+\sin(\phi)\ket{1})\otimes\ket{1}^{\otimes\mu}\otimes\ket{0})\\
     &=P(\cos(\phi)(\ket{0}\otimes\ket{1}^{\otimes\mu}\otimes\ket{0})+\sin(\phi)(\ket{1}\otimes\ket{1}^{\otimes\mu}\otimes\ket{0})),
     \end{aligned}
     \end{equation}
     Let $\ket{\psi_0}=\ket{0}\otimes\ket{1}^{\otimes\mu}$ and $\ket{\psi_1}=\ket{1}\otimes\ket{1}^{\otimes\mu}= \ket{1}^{\otimes\mu+1}$,
     \begin{equation}
     \begin{aligned}
     P(\ket{\psi}\otimes\ket{0})=P(\cos(\phi)(\ket{\psi_0}\otimes\ket{0})+\sin(\phi)(\ket{\psi_1}\otimes\ket{0})),
     \end{aligned}
     \end{equation} 
     then, the result of applying the operator $P$ on the system $(\cos(\phi)(\ket{\psi_0}\otimes\ket{0})+\sin(\phi)(\ket{\psi_1}\otimes\ket{0}))$, is given by the following equation,
     
     \begin{equation}
     \begin{aligned}
     P(\cos(\phi)(\ket{\psi_0}\otimes\ket{0})+\sin(\phi)(\ket{\psi_1}\otimes\ket{0}))=
     \cos(\phi)(\ket{\psi_0}\otimes(\frac{1+k^{d_0}}{2}\ket{0} + \frac{1-k^{d_0}}{2}\ket{1}))+\\ \sin(\phi)(\ket{\psi_1}\otimes(\frac{1+k^{d_1}}{2}\ket{0} + \frac{1-k^{d_1}}{2}\ket{1})),
     \end{aligned}
     \end{equation}
     where, $d_0=\mu$ and $d_1=\mu+1$ which are the 1-density of $\ket{\psi_0}$ and $\ket{\psi_1}$ respectively. 
     
     After Applying $P$ on the system, the probabilities of finding the auxiliary qubit $\ket{ax}$ in state $\ket{0}$ or $\ket{1}$ when measured, can be calculated by getting the square of the amplitudes in equation (13) as shown below,
     \begin{equation}
     \begin{aligned}
     Pr_0(\ket{ax}=\ket{0}) = \abs{\cos(\phi)(\frac{1+k^{d_0}}{2})}^2 + \abs{\sin(\phi)(\frac{1+k^{d_1}}{2})}^2 
     \end{aligned}
     \end{equation}
     since, $\frac{1+k^{d_0}}{2} = cos{(\frac{d_0\pi}{2t})}$ and $\frac{1+k^{d_1}}{2} = cos{(\frac{d_1\pi}{2t})}$ from equation (10)
     therefore,
     \begin{equation}
     \begin{aligned}
     Pr_0(\ket{ax}=\ket{0})=\cos^2(\phi)\cos^2{(\frac{d_0\pi}{2t})}+\sin^2(\phi)\cos^2{(\frac{d_1\pi}{2t})}, 
     \end{aligned}
     \end{equation}
     similarly for $	Pr_0(\ket{ax}=\ket{1})$
     since, $\frac{1-k^{d_0}}{2} = sin{(\frac{d_0\pi}{2t})}$ and $\frac{1-k^{d_1}}{2} = sin{(\frac{d_1\pi}{2t})}$ from equation (11)
     therefore,
     \begin{equation}
     \begin{aligned}
     Pr_0(\ket{ax}=\ket{1})=\cos^2(\phi)\sin^2{(\frac{d_0\pi}{2t})}+\sin^2(\phi)\sin^2{(\frac{d_1\pi}{2t})},
     \end{aligned}
     \end{equation}
     given that $t=2\mu+1$.
     
     A sharp measurement is applied on the auxiliary qubit $\ket{ax}$ for an arbitrary number of $j$-iterations causing the superposition of the unknown qubit $\ket{\psi}$ to move in a random walk manner. At each iteration if $\ket{ax}$ is found to be $\ket{0}$ then, the amplitudes $\alpha_{j+1}$ and $\beta_{j+1}$ of the unknown qubit will be updated as follows,      
     \begin{equation}
     \alpha_{j+1}=\frac{\alpha_j\cos(\frac{d_0\pi}{2t})}{\sqrt{pr_j(\ket{ax}=\ket{0})}}, 
     \beta_{j+1}=\frac{\beta_j\cos(\frac{d_1\pi}{2t})}{\sqrt{pr_j(\ket{ax}=\ket{0})}}.
     \end{equation} 
     If $\ket{ax}$ is found to be $\ket{1}$ then, the amplitudes will be updated as follows,
     \begin{equation}
     \alpha_{j+1}=\frac{\alpha_j\sin(\frac{d_0\pi}{2t})}{\sqrt{pr_j(\ket{ax}=\ket{1})}}, 
     \beta_{j+1}=\frac{\beta_j\sin(\frac{d_1\pi}{2t})}{\sqrt{pr_j(\ket{ax}=\ket{1})}},
     \end{equation}
     where $\alpha_0=\cos(\phi)$ and $\beta_0=\sin(\phi)$ and the probabilities of finding $\ket{ax}$ in state $\ket{0}$ or $\ket{1}$ when measured will be shown as,
     \begin{equation}
     Pr_j(\ket{ax}=\ket{0})=\alpha^2_j\cos^2(\frac{d_1\pi}{2t})+\beta^2_j\cos^2(\frac{d_0 \pi}{2t}),
     \end{equation} 
     \begin{equation}
     Pr_j(\ket{ax}=\ket{1})=\alpha^2_j\sin^2(\frac{d_1\pi}{2t})+\beta^2_j\sin^2(\frac{d_0\pi}{2t}).
     \end{equation}
     
    \section{The Proposed Algorithm}
     In the proposed algorithm, the measurement-based quantum random walk approach is used, such that a register is prepared by appending an unknown qubit $\ket{\psi}$ to an auxiliary qubit $\ket{ax}$, with an operator $P$ being applied on the register $(\ket{\psi}\otimes\ket{ax})$. The algorithm is then iterated for an arbitrary number of iterations $j=r$, where at each iteration the auxiliary qubit $\ket{ax}$ is measured such that it can be found in state $\ket{0}$ or $\ket{1}$ with probabilities $Pr_j(\ket{ax}=\ket{0})$ and $Pr_j(\ket{ax}=\ket{1})$ as in equations (20) and (21) respectively. After $j$-iterations the state of the qubit will be determined approximately by using two counters $j_0$ and $j_1$, where $j_0+j_1=j$ if $j_0$ is greater than $j_1$ then, $\ket{\psi}$ is in state $\ket{0}$ or $\ket{+}$ otherwise it is in state $\ket{1}$ or $\ket{-}$. At each iteration, if the measurement outcome of $\ket{ax}$ is found to be $\ket{0}$, then the amplitudes $\alpha$ and $\beta$ are updated as in equation (18) and a counter $j_0$ is incremented, otherwise if $\ket{ax}$ is found to be $\ket{1}$, then the amplitudes are updated as in equation (19) and a counter $j_1$ is incremented. At each iteration after incrementing $j_0$ or $j_1$, an approximate value of the amplitudes $\alpha_{approx}$ and $\beta_{approx}$ will be calculated as follows,
     \begin{equation}
     \alpha_{approx}=\frac{j_0}{j_0+j_1}, \beta_{approx}=\frac{j_1}{j_0+j_1}.
     \end{equation}
     A decision is then taken after a number of iterations $k$, where $k \leqslant r$ to determine if the unknown qubit $\ket{\psi}$ lies in the Hadamard or the computational basis, where a condition is applied to check if $\alpha_{approx}$ lies within an interval $[I_1,I_2]$. If $\alpha_{approx}$ is found to lie within that interval, then $\ket{\psi}$ is decided to be in the Hadamard basis, and so apply a Hadamard gate ($H$-gate), where the values of the amplitudes will be updated as follows,
     \begin{equation}
     \alpha_{j+1}=\frac{\alpha_j+\beta_j}{\sqrt{2}},\beta_{j+1}=\frac{\alpha_j-\beta_j}{\sqrt{2}},
     \end{equation}
     otherwise, $\ket{\psi}$ will be decided to be in the computational basis. Finally, to approximately check the state of $\ket{\psi}$, if it is in the Hadamard basis and $j_1$ is greater than $j_0$, then it is closer to $\ket{-}$, otherwise it is closer to $\ket{+}$, and if it is in the computational basis then, it is closer to $\ket{1}$ if $j_1$ is greater than $j_0$, otherwise it is closer to $\ket{0}$. 
     
     \begin{breakablealgorithm}
     	\caption{Quantum State Discrimination Algorithm}
     	
     	\begin{algorithmic}[1]
     		
     		\noindent{Given an unknown qubit $\ket{\psi}$}
     		\State  Initialize $\ket{ax}$ in state $\ket{0}$ 
     		\State  Prepare $(\ket{\psi}\otimes\ket{ax})$
     		\State $j_0=0$, $j_1=0$
     		\State Apply operator $P$ on the system $(\ket{\psi}\otimes\ket{ax})$
     		\LeftComment*{Iterate for arbitrary number of iterations $r$}
     		\For{$j=1$ to $r$} 
     		\State {Measure  $\ket{ax}$}
     		\If  {$\ket{ax}=\ket{0}$}
     		\State  $\alpha_j$ and $\beta_j$ will be updated as in equation (18)
     		\State Increment $j_0$
     		\Else
     		\State  $\alpha_j$ and $\beta_j$ will be updated as in equation (19)
     		\State Increment $j_1$
     		\EndIf
     		\State Reset $\ket{ax}$ to state $\ket{0}$
     		\State Calculate $\alpha_{approx}$ and $\beta_{approx}$ as in equation (20)
     		\LeftComment*{To Check if $\ket{\psi}$ lies in the Hadamard basis:}
     		\newline
     		\Comment{Take a decision after a certain number of iterations $k$, $k \leqslant r$}
     		\If {$j=k$}
     		\LeftComment{Check if $\alpha_{approx}$ lies within an interval $[I_1,I_2]$}
     		\If {$I_1<\alpha_{approx}<I_2$} 
     		\LeftComment{$\ket{\psi}$ is decided to lie in the Hadamard basis.}
     		\State Apply $H$-gate and so, $\alpha_j$ and $\beta_j$ will be updated as in equation (23).
     		\EndIf
     		\EndIf		
     		\EndFor
     		
     		\If {$\ket{\psi}$ is in the Hadamard basis}
     		\If {$j_1 > j_0$}
     		\State $\ket{\psi}$ is closer to $\ket{-}$
     		\Else
     		\State $\ket{\psi}$ is closer to $\ket{+}$
     		\EndIf
     		
     		\Else
     		\If {$j_1 > j_0$}
     		\State $\ket{\psi}$ is closer to $\ket{1}$
     		\Else
     		\State $\ket{\psi}$ is closer to $\ket{0}$
     		\EndIf
     		\EndIf	
     	\end{algorithmic}
     \end{breakablealgorithm}
 
     \section{Results and Discussion}
     
     The results of determining approximately the state of the unknown qubit $\ket{\psi}$ are obtained using the quantum state discrimination algorithm employed by weak measurement and partial negation operator. To achieve the aim of the proposed approach, the algorithm has been tested for $10^{3}$ trials for each of the cases $\ket{\psi}=\ket{0}$, $\ket{\psi}=\ket{1}$, $\ket{\psi}=\ket{+}$ and $\ket{\psi}=\ket{-}$, where at each iteration the algorithm uses $r=100$ iterations as shown in figure(4). It can be revealed from figures (4a) and (4b) that ${\alpha_{approx}}$ is closer to 1, as a result $\ket{\psi}$ is either in state $\ket{0}$ or $\ket{+}$. Similarly for the cases in figures (4c) and (4d), ${\alpha_{approx}}$ is found to be closer to 0, hence  $\ket{\psi}$ is either in state $\ket{1}$ or $\ket{-}$. Note that, applying $H$-gate on $\ket{\psi}$, when $\ket{\psi}=\ket{+}$ or $\ket{\psi}=\ket{-}$  will transform the qubit to the state $\ket{0}$ or $\ket{1}$ respectively.  
     \begin{center}
     	\begin{tabular}{ c c } 
     		
     	\textbf{(a)} & \textbf{(b)} \\
     	
     	\textbf{$\ket{\psi}=\ket{0}$} & \textbf{$\ket{\psi}=\ket{+}$} \\ 
     	\includegraphics[scale=0.4]{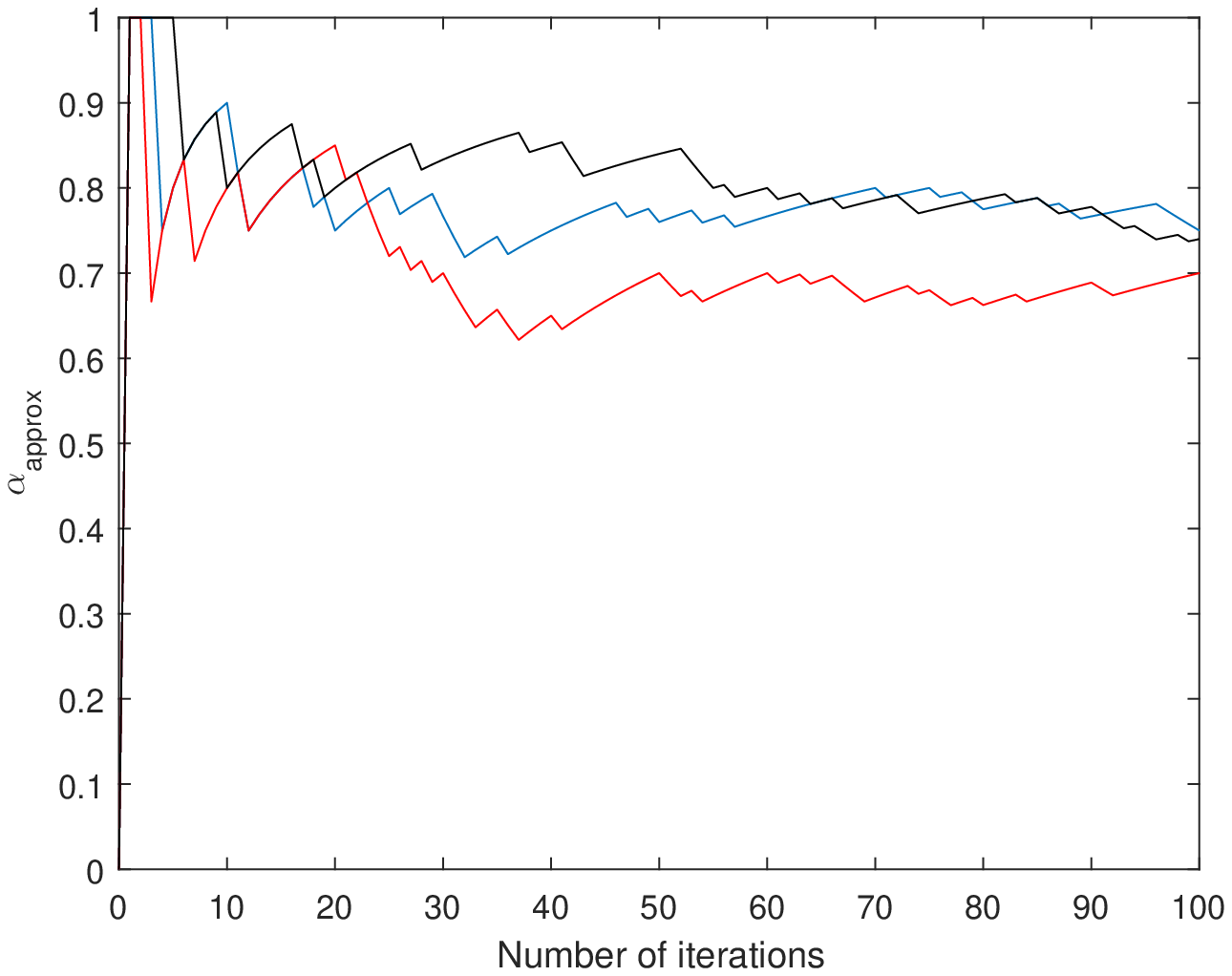} & \includegraphics[scale=0.4]{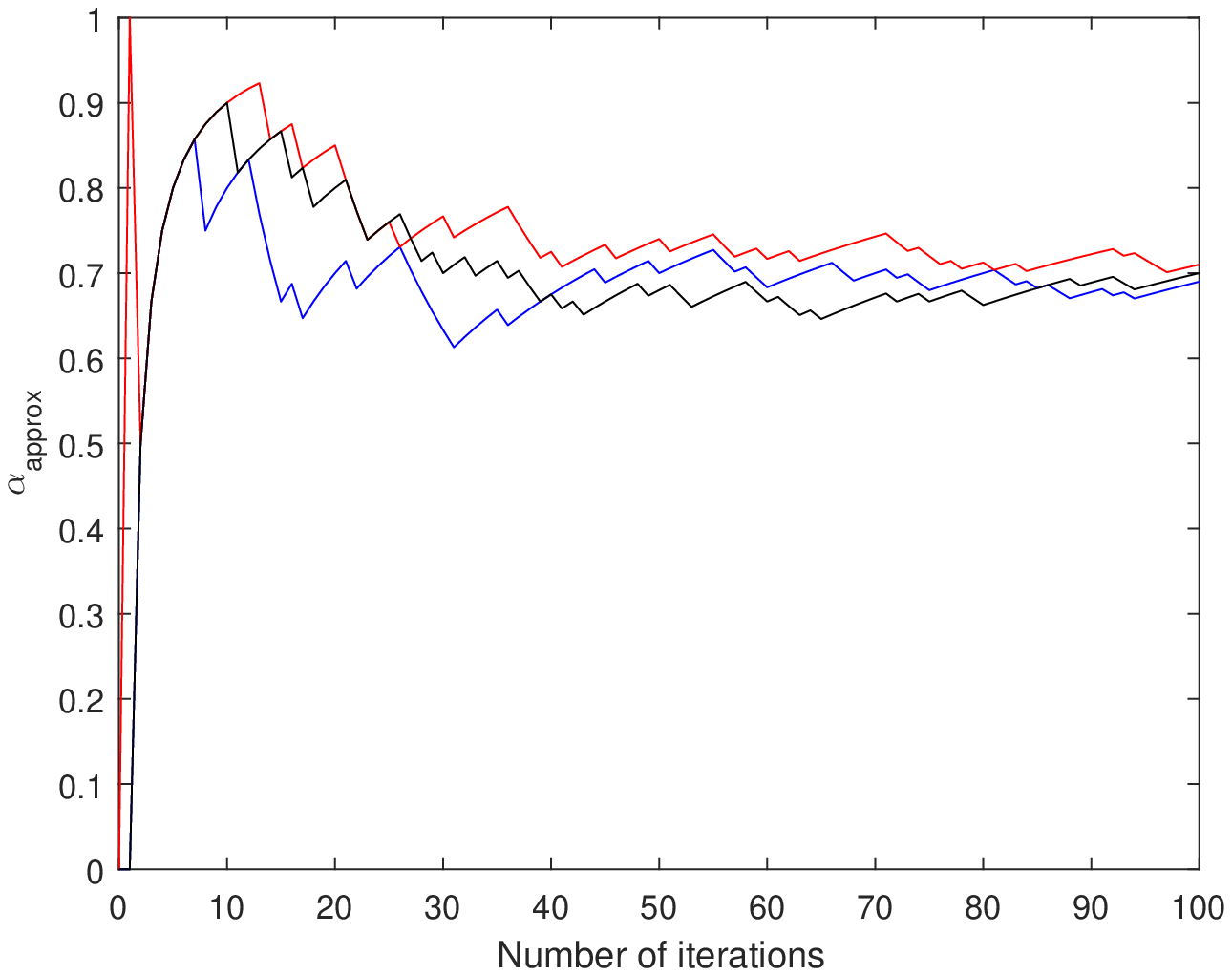} \\ 
     	\textbf{(c)} & \textbf{(d)} \\ 
     	\textbf{$\ket{\psi}=\ket{1}$} & \textbf{$\ket{\psi}=\ket{-}$} \\
     	\includegraphics[scale=0.4]{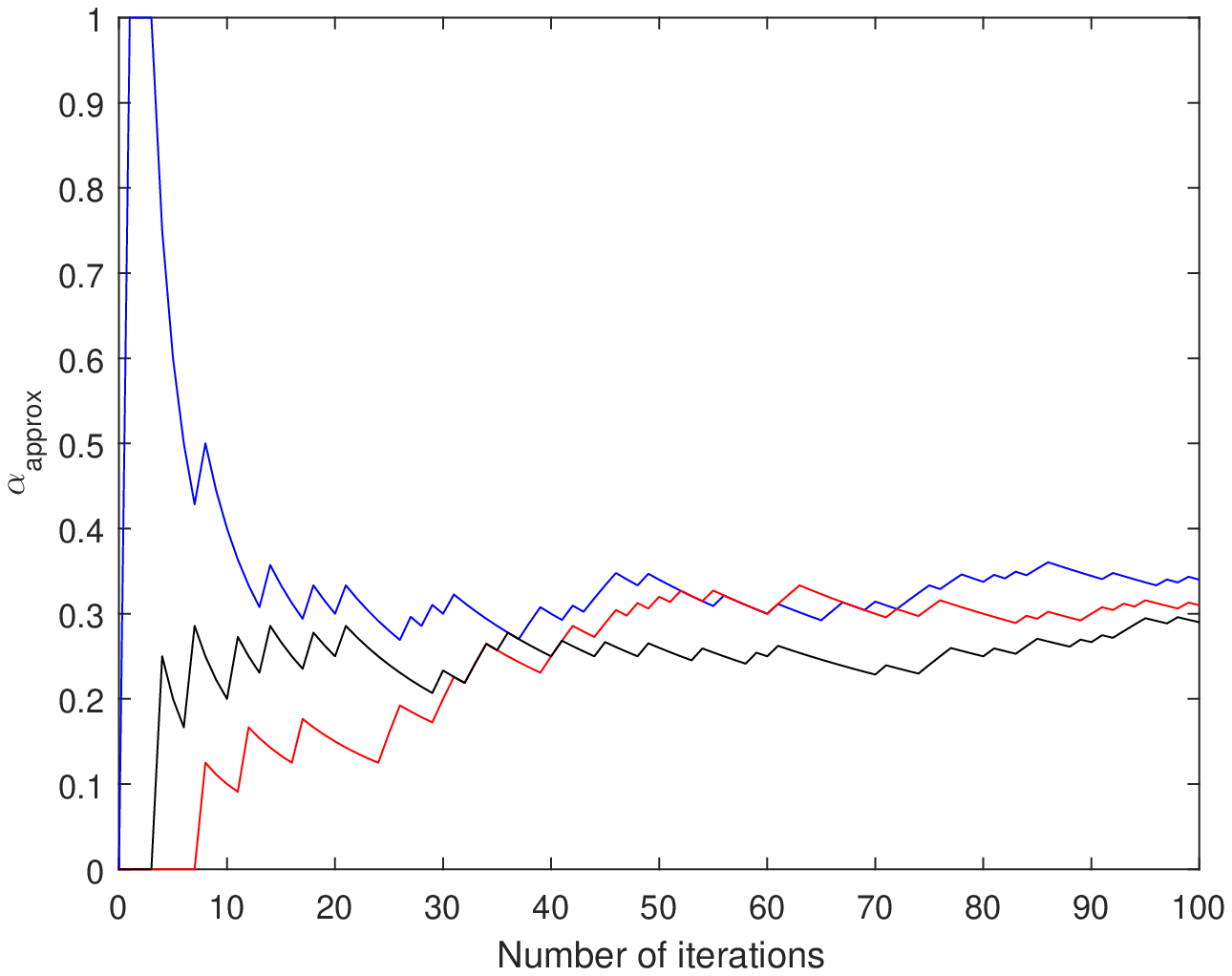} & \includegraphics[scale=0.4]{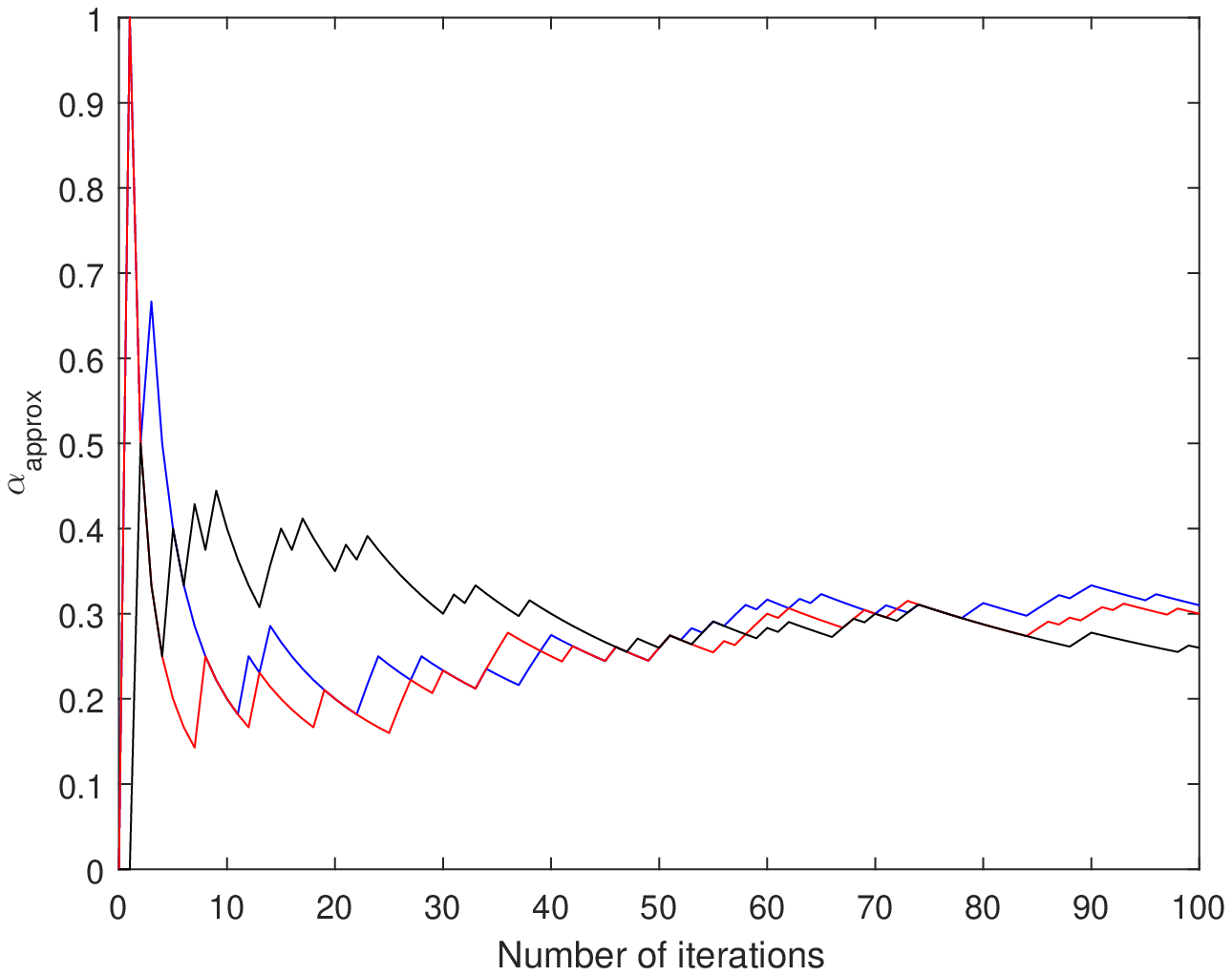}

     	\end{tabular}
     	\captionof{figure}{Three sample runs for the random walks for each of the considered cases, $\ket{\psi}=\ket{0}$, $\ket{\psi}=\ket{+}$, $\ket{\psi}=\ket{1}$ and $\ket{\psi}=\ket{-}$.}
     \end{center}             
     
     Based on the values of ${\alpha_{approx}}$, the decision of applying $H$-gate on the unknown qubit will be taken i.e. changing the measurement to the Hadamard basis. It was found that the best results are obtained when taking the decision after 2 iterations ($j=k=2$), were ${\alpha_{approx}}$ takes the values 0 or 1 or 0.5 in a probabilistic manner for each iteration. For $k>2$, the values of ${\alpha_{approx}}$ lies in range $[0,1]$.    
     
     To verify the proposed algorithm, $\ket{\psi}$ is assumed to be in the computational basis as in \cite{Younes2017}, where the interval [$I_1$,$I_2$] defined in Section 5, is $]0,1[-0.5$, such that $H$-gate will not be applied within these intervals. The obtained results were compatible with the results in \cite{Younes2017}, i.e. $\ket{\psi}=\ket{0}$ and $\ket{\psi}=\ket{1}$ are determined with certainty with number of dummy qubits $\mu=1$. However, for the cases $\ket{\psi}=\ket{+}$ and $\ket{\psi}=\ket{-}$, $\ket{\psi}$ is found to be $\ket{0}$ or $\ket{1}$ with probability ranging from 45\% to 55\%.
     
     Applying $H$-gate on all of the considered cases, i.e. assuming that  $\ket{\psi}$ is in the Hadamard basis, where the interval $[I_1,I_2]=[0,1]$, the algorithm determines the state of $\ket{\psi}$ with approximately $95\%$ for $\ket{\psi}=\ket{+}$ and $\ket{\psi}=\ket{-}$. In this case the state of $\ket{\psi}$ is not determined with certainty, since $H$-gate is applied on $\ket{\psi}$ using equation (23) after two iterations, i.e. on the updated values of the amplitudes $\alpha$ and $\beta$ which are calculated using equations (18) or (19). On the other hand, for the  cases $\ket{\psi}=\ket{0}$ and $\ket{\psi}=\ket{1}$, the algorithm gives probability ranging from 46\% to 54\% for $\ket{\psi}$ to be found in state $\ket{0}$ or $\ket{1}$.
     
     To satisfy the purpose of the proposed algorithm with the highest percentage of success, it was found that the best range for the interval $[I_1,I_2]=]0,1[$ i.e. $H$-gate will be applied only at ${\alpha_{approx}}=0.5$. Given that the value of ${\alpha_{approx}}$ is computed when it is in the computational basis i.e. before applying $H$-gate. 
     
     \begin{center}
     	\includegraphics[scale=0.4]{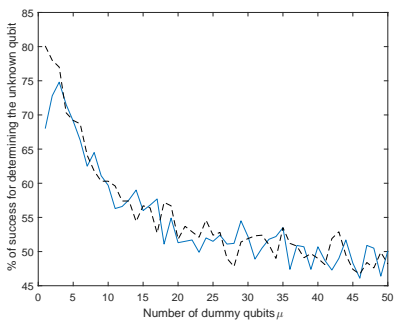}
     	\captionof{figure}{Percentages of success for determining the state of $\ket{\psi}$ for different values of $\mu$, where $\mu \in [1,50] $. The scattered line represents the cases $\ket{\psi}=\ket{0}$ and $\ket{\psi}=\ket{1}$, while the solid line represents the cases $\ket{\psi}=\ket{+}$ and $\ket{\psi}=\ket{-}$.}
     \end{center}
  To maximize the success rate for all of the considered cases, the number of dummy qubits $\mu$ is set to 2. It can be revealed form figure (5), that for $\mu>2$, the percentage of success for determining the state of $\ket{\psi}$ decreases, since adding more dummy qubits to the system, minimizes the strength of the weak measurement, and hence  $\ket{\psi}$, will collapse to either  $\ket{0}$ or $\ket{1}$ with less probability \cite{Younes2017}.
     \begin{center}
     		\begin{table}[hbt!]
     			\setlength{\tabcolsep}{0.5pt}
     			\footnotesize
     			\begin{tabular}{lllllllll}
     				\textbf{} &  &  &  & \textbf{Table 2a} &  &  &  &  \\ \hline
     				\multicolumn{1}{|l|}{} & \multicolumn{4}{c|}{$\ket{\psi}=\ket{0}$} & \multicolumn{4}{c|}{$\ket{\psi}=\ket{+}$} \\ \hline
     				\multicolumn{1}{|l|}{Actual Basis} & \multicolumn{4}{c|}{Computational basis} & \multicolumn{4}{c|}{Hadamard basis} \\ \hline
     				\multicolumn{1}{|l|}{\begin{tabular}[c]{@{}l@{}} Assumed Basis\end{tabular}} & \multicolumn{2}{c|}{\begin{tabular}[c]{@{}c@{}}Hadamard\\ basis\end{tabular}} & \multicolumn{2}{c|}{\begin{tabular}[c]{@{}c@{}}Computational\\ basis\end{tabular}} & \multicolumn{2}{c|}{\begin{tabular}[c]{@{}c@{}}Hadamard\\ basis\end{tabular}} & \multicolumn{2}{c|}{\begin{tabular}[c]{@{}c@{}}Computational\\ basis\end{tabular}} \\ \hline
     				\multicolumn{1}{|c|}{\begin{tabular}[c]{@{}c@{}}Transformation \\ applied\end{tabular}}
     				& \multicolumn{2}{c|}{$H\ket{\psi}$} & 
     				\multicolumn{2}{c|}{No transformation} 
     				& \multicolumn{2}{c|}{$H\ket{\psi}$} & 
     				\multicolumn{2}{c|}{No transformation} \\ \hline
     				\multicolumn{1}{|l|}{\begin{tabular}[c]{@{}l@{}}\% of trials\\ transformation\\is applied based\\on $\alpha_{approx}$.\end{tabular}} & 
     				\multicolumn{2}{c|}{$\simeq 44.953\%$} & 
     				\multicolumn{2}{c|}{$\simeq 55.047\%$} & 
     				\multicolumn{2}{c|}{$\simeq 45.498\%$} & 
     				\multicolumn{2}{c|}{$\simeq 54.502\%$} \\ \hline
     				\multicolumn{1}{|l|}{\begin{tabular}[c]{@{}l@{}}Correctness for\\ determining $\ket{\psi}$. \end{tabular}} & \multicolumn{1}{c|}{\begin{tabular}[c]{@{}c@{}}Success: \\ $\simeq 22.645\%$\end{tabular}} & \multicolumn{1}{c|}{\begin{tabular}[c]{@{}c@{}}Failure: \\ $\simeq 22.226\%$\end{tabular}} & \multicolumn{1}{c|}{\begin{tabular}[c]{@{}c@{}}Success: \\ $\simeq 55.047\%$\end{tabular}} & \multicolumn{1}{c|}{\begin{tabular}[c]{@{}c@{}}Failure: \\ $\simeq 0.029\%$\end{tabular}} & \multicolumn{1}{c|}{\begin{tabular}[c]{@{}c@{}}Success: \\ $\simeq 45.498\%$\end{tabular}} & \multicolumn{1}{c|}{\begin{tabular}[c]{@{}c@{}}Failure: \\ $\simeq 0.026\%$\end{tabular}} & \multicolumn{1}{c|}{\begin{tabular}[c]{@{}c@{}}Success:\\  $\simeq 27.192\%$\end{tabular}} & \multicolumn{1}{c|}{\begin{tabular}[c]{@{}c@{}}Failure: \\ $\simeq 27.24\%$\end{tabular}} \\ \hline
     				\multicolumn{1}{|l|}{\begin{tabular}[c]{@{}l@{}}Total \% of \\ success.\end{tabular}} 
     				& \multicolumn{4}{c|}{$\simeq 77.656\%$} 
     				& \multicolumn{4}{c|}{$\simeq 72.32\%$} \\ \hline
     				&  &  & \textbf{} & \textbf{} &  &  &  &  \\
     				&  &  &  & \textbf{Table 2b} &  &  &  &  \\ \hline
     				\multicolumn{1}{|l|}{} & \multicolumn{4}{c|}{$\ket{\psi}=\ket{1}$} & \multicolumn{4}{c|}{$\ket{\psi}=\ket{-}$} \\ \hline
     				\multicolumn{1}{|l|}{Actual Basis} & \multicolumn{4}{c|}{Computational basis} & \multicolumn{4}{c|}{Hadamard basis} \\ \hline
     				\multicolumn{1}{|l|}{\begin{tabular}[c]{@{}l@{}} Assumed Basis\end{tabular}} & \multicolumn{2}{c|}{\begin{tabular}[c]{@{}c@{}}Hadamard\\ basis\end{tabular}} & \multicolumn{2}{c|}{\begin{tabular}[c]{@{}c@{}}Computational\\ basis\end{tabular}} & \multicolumn{2}{c|}{\begin{tabular}[c]{@{}c@{}}Hadamard\\ basis\end{tabular}} & \multicolumn{2}{c|}{\begin{tabular}[c]{@{}c@{}}Computational\\ basis\end{tabular}} \\ \hline
     				\multicolumn{1}{|c|}{\begin{tabular}[c]{@{}c@{}}Transformation \\ applied\end{tabular}}
     				& \multicolumn{2}{c|}{$H\ket{\psi}$} & 
     				\multicolumn{2}{c|}{No transformation} 
     				& \multicolumn{2}{c|}{$H\ket{\psi}$} & 
     				\multicolumn{2}{c|}{No transformation} \\ \hline
     				\multicolumn{1}{|l|}{\begin{tabular}[c]{@{}l@{}}\% of trials\\ transformation\\is applied based\\on $\alpha_{approx}$.\end{tabular}} & 
     				\multicolumn{2}{c|}{$\simeq 45.06\%$} & 
     				\multicolumn{2}{c|}{$\simeq 54.94\%$} & 
     				\multicolumn{2}{c|}{$\simeq 45.2\%$} & 
     				\multicolumn{2}{c|}{$\simeq 54.801\%$} \\ \hline
     				\multicolumn{1}{|l|}{\begin{tabular}[c]{@{}l@{}}Correctness for\\ determining $\ket{\psi}$ \end{tabular}} & \multicolumn{1}{c|}{\begin{tabular}[c]{@{}c@{}}Success: \\ $\simeq 22.526\%$\end{tabular}} & \multicolumn{1}{c|}{\begin{tabular}[c]{@{}c@{}}Failure:\\  $\simeq 22.464\%$\end{tabular}} & \multicolumn{1}{c|}{\begin{tabular}[c]{@{}c@{}}Success: \\ $\simeq 54.936\%$\end{tabular}} & \multicolumn{1}{c|}{\begin{tabular}[c]{@{}c@{}}Failure: \\ $\simeq 0.021\%$\end{tabular}} & \multicolumn{1}{c|}{\begin{tabular}[c]{@{}c@{}}Success: \\ $\simeq 45.199\%$\end{tabular}} & \multicolumn{1}{c|}{\begin{tabular}[c]{@{}c@{}}Failure:\\  $\simeq 0.03\%$\end{tabular}} & \multicolumn{1}{c|}{\begin{tabular}[c]{@{}c@{}}Success: \\ $\simeq  27.408\%$\end{tabular}} & \multicolumn{1}{c|}{\begin{tabular}[c]{@{}c@{}}Failure: \\ $\simeq  27.324\%$\end{tabular}} \\ \hline
     				\multicolumn{1}{|l|}{\begin{tabular}[c]{@{}l@{}}Total \% of\\ success.\end{tabular}}
     				 & \multicolumn{4}{c|}{$\simeq 77.462\%$} 
     				 & \multicolumn{4}{c|}{$\simeq 72.607\%$} \\ \hline
     			\end{tabular}
     		\end{table}
     	\captionof{table} {The percentages of success and failure after $10^{3}$ trials for determining the state of $\ket{\psi}$, when it is assumed to be in the Hadamard or the computational basis, in addition to the total perecentage of success for determining $\ket{\psi}$ when $\ket{\psi}=\ket{0}$, $\ket{\psi}=\ket{+}$ represented in Table 2a and $\ket{\psi}=\ket{1}$, $\ket{\psi}=\ket{+}$ represented in Table 2b.}
     \end{center}
          Table 2 illustrates the percentages of success and failure for determining the state of $\ket{\psi}$ when it is measured in the Hadamard or the computational basis after $10{^3}$ trials. For each trial the algorithm runs for $r=100$ iterations, and $\ket{\psi}$ is assumed to be encoded in the Hadamard or the computational basis. Given that, the initial assumption in the proposed algorithm is that $\ket{\psi}$ lies in the computational basis.
      
          Table 2a represents the cases  $\ket{\psi}=\ket{0}$ and  $\ket{\psi}=\ket{+}$. It can be seen that in case of $\ket{\psi}=\ket{0}$, $\ket{\psi}$ is assumed to be encoded the computational basis with approximately $55.047\%$ of the total trials, which is considered to be the correct assumption such that no transformation will be applied and the state of $\ket{\psi}$ is determined with approximately $55.047\%$ success. However, the percentage of trials $\ket{\psi}$ is assumed to be encoded in the Hadamard basis where $H$-gate is applied is approximately $44.953.5\%$ where, $\ket{\psi}$ is found to be in state  $\ket{0}$ or  $\ket{1}$ with a 50\% probability when measured, resulting in approximately $22.645\%$ success. Therefore, the total percentage of success is approximately $77.656\%$, which is the sum of the probabilities of success when $\ket{\psi}$ is measured in the Hadamard or the computational basis.  On the contrary for $\ket{\psi}=\ket{+}$, it can be revealed that, the percentage of assuming $\ket{\psi}$ is encoded in the Hadamard basis is approximately $45.498\%$ of the total number of trials which is considered as the correct assumption, and hence $H$-gate is applied on $\ket{\psi}$, accordingly the state of $\ket{\psi}$ is determined with approximately $45.200\%$ accuracy. on the other hand, approximately $54.801\%$ of the total trials assume that $\ket{\psi}$ is encoded in the computational basis, where no transformation is applied, and hence $\ket{\psi}$ will be found in state $\ket{0}$ or $\ket{1}$ with $50\%$ probability, resulting in approximately $27.192\%$ success. As a result, the total percentage of success for determining $\ket{\psi}$ is approximately 72.320\%.   
          
          Similarly in Table 2b, representing the cases $\ket{\psi}=\ket{1}$ and  $\ket{\psi}=\ket{-}$. It can be seen that for $\ket{\psi}=\ket{1}$, the percentage of trials no transformation is being applied on $\ket{\psi}$ is approximately $54.940\%$, where the state of $\ket{\psi}$ is determined correctly with approximately $54.936\%$ success. However, the percentage of trials no transformation is being applied is approximately $45.060\%$ which is in this case an incorrect decision where $\ket{\psi}$ is determined with a percentage of success approximately $22.526\%$. Hence, the total percentage of success is approximately $77.462\%$ accuracy. In case of $\ket{\psi}=\ket{-}$, it can be seen from the table that the percentage of trials a transformation is being applied is approximately $45.2\%$, which is considered as the correct decision, resulting in approximately $45.199\%$ success for determining $\ket{\psi}$. The percentage of taking an incorrect decision i.e. not applying the transformation on $\ket{\psi}$ is approximately $54.801\%$, where $\ket{\psi}$ is determined with approximately $27.408\%$ success. Accordingly, the total percentage of success is approximately $72.607\%$ accuracy.
          
           The probability distribution of the the success and failure rates of the proposed algorithm for determining the state of $\ket{\psi}$ can be illustrated in figure (6). The figure shows 100 random walks for the cases $\ket{\psi}=\ket{0}$, $\ket{\psi}=\ket{1}$, $\ket{\psi}=\ket{+}$ and $\ket{\psi}=\ket{-}$. In Figures (6a) and (6b) representing the case $\ket{\psi}=\ket{0}$ and $\ket{\psi}=\ket{+}$ respectively, it can be revealed that $\alpha_{approx}$ is closer to 1 with a high probability. However, for the cases $\ket{\psi}=\ket{1}$, $\ket{\psi}=\ket{-}$ demonstrated in figures (6c) and (6d), $\alpha_{approx}$ is biased towards 0 with higher probability.  
           
          \begin{center}
          	\begin{tabular}{ c c  } 
          	\textbf{(a)} & \textbf{(b)} \\
          	\textbf{$\ket{\psi}=\ket{0}$} & \textbf{$\ket{\psi}=\ket{+}$} \\ 
          	\includegraphics[scale=0.4]{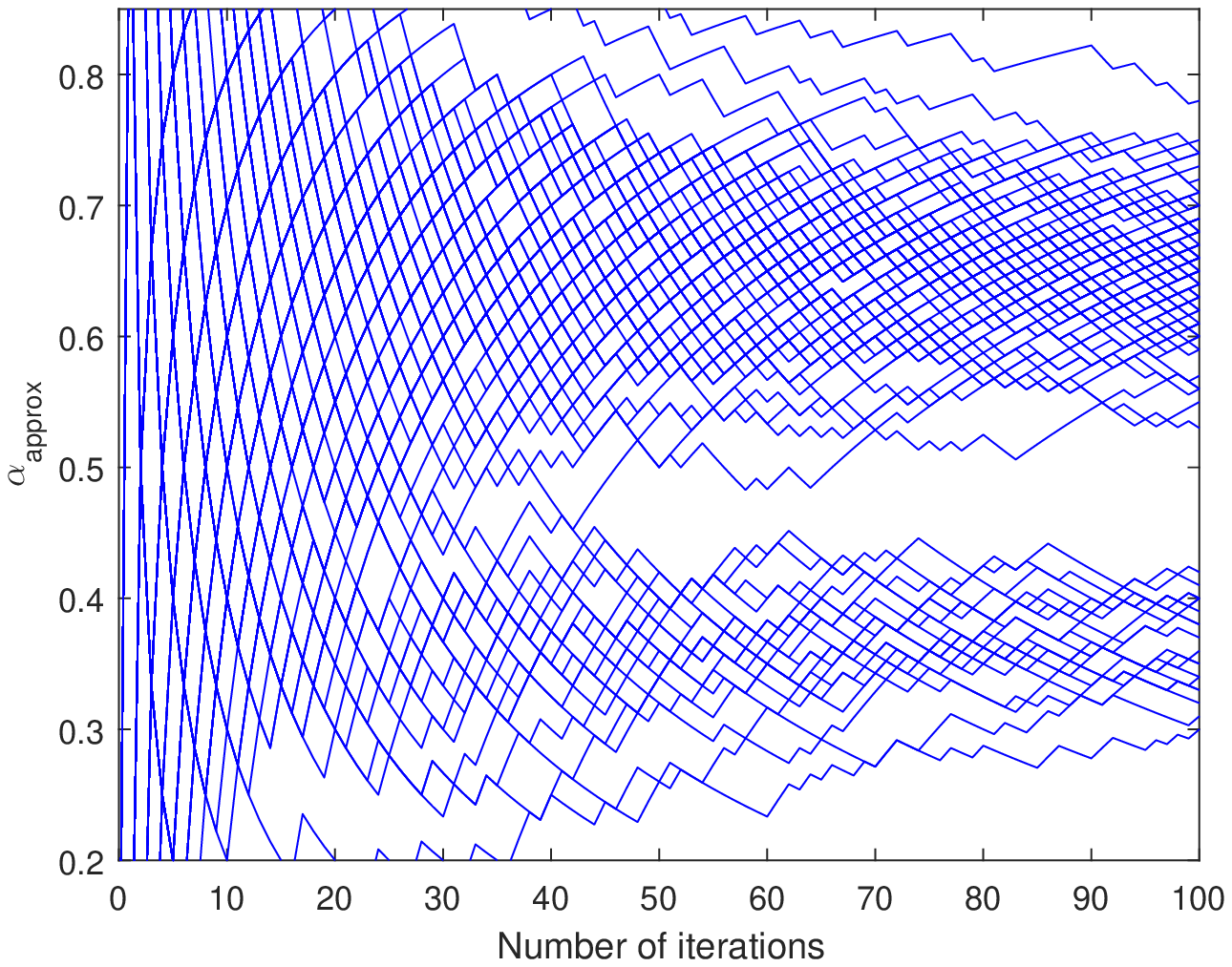} & \includegraphics[scale=0.4]{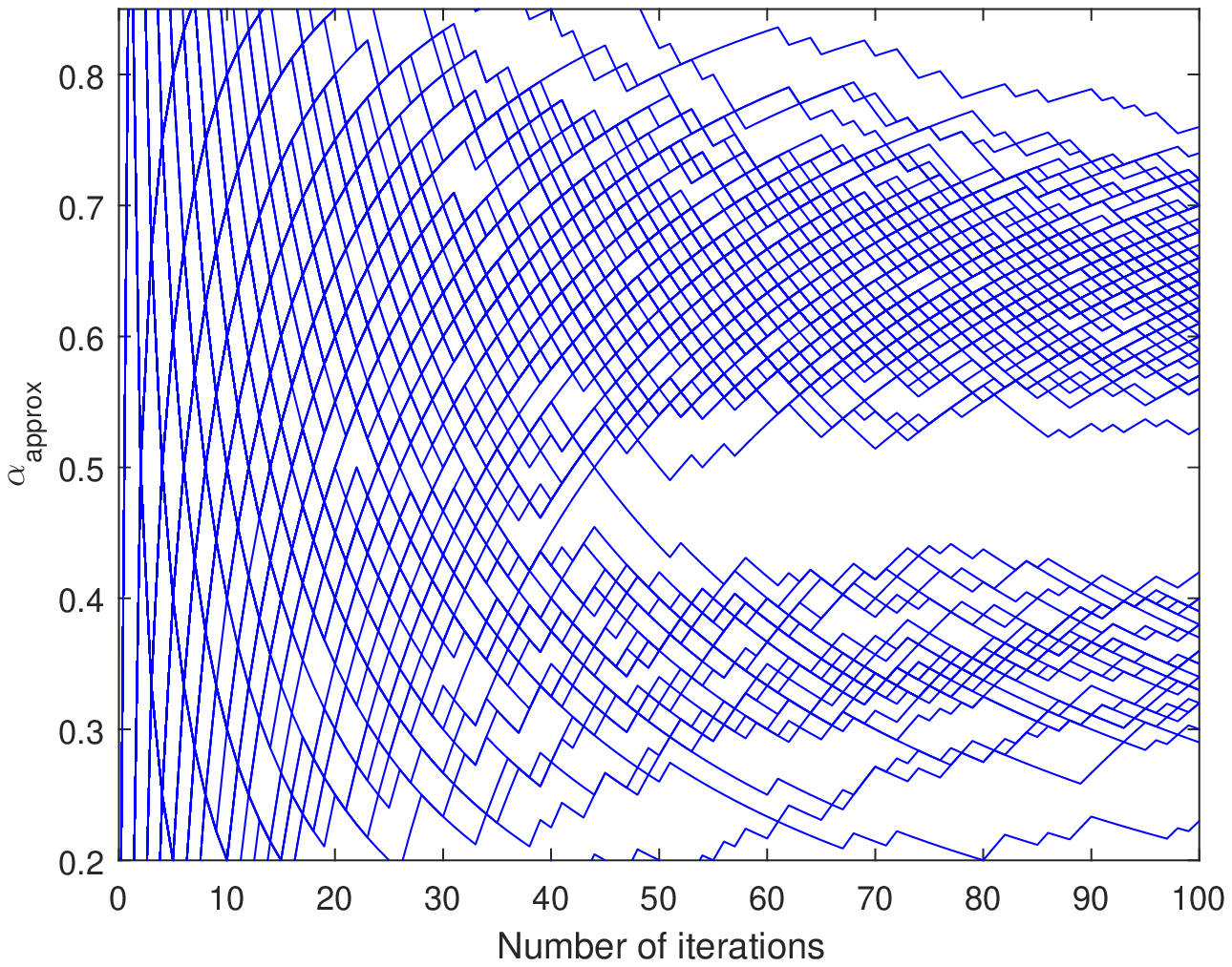} \\ 
          	\textbf{(c)} & \textbf{(d)} \\ 
          	\textbf{$\ket{\psi}=\ket{1}$} & \textbf{$\ket{\psi}=\ket{-}$} \\
          	\includegraphics[scale=0.4]{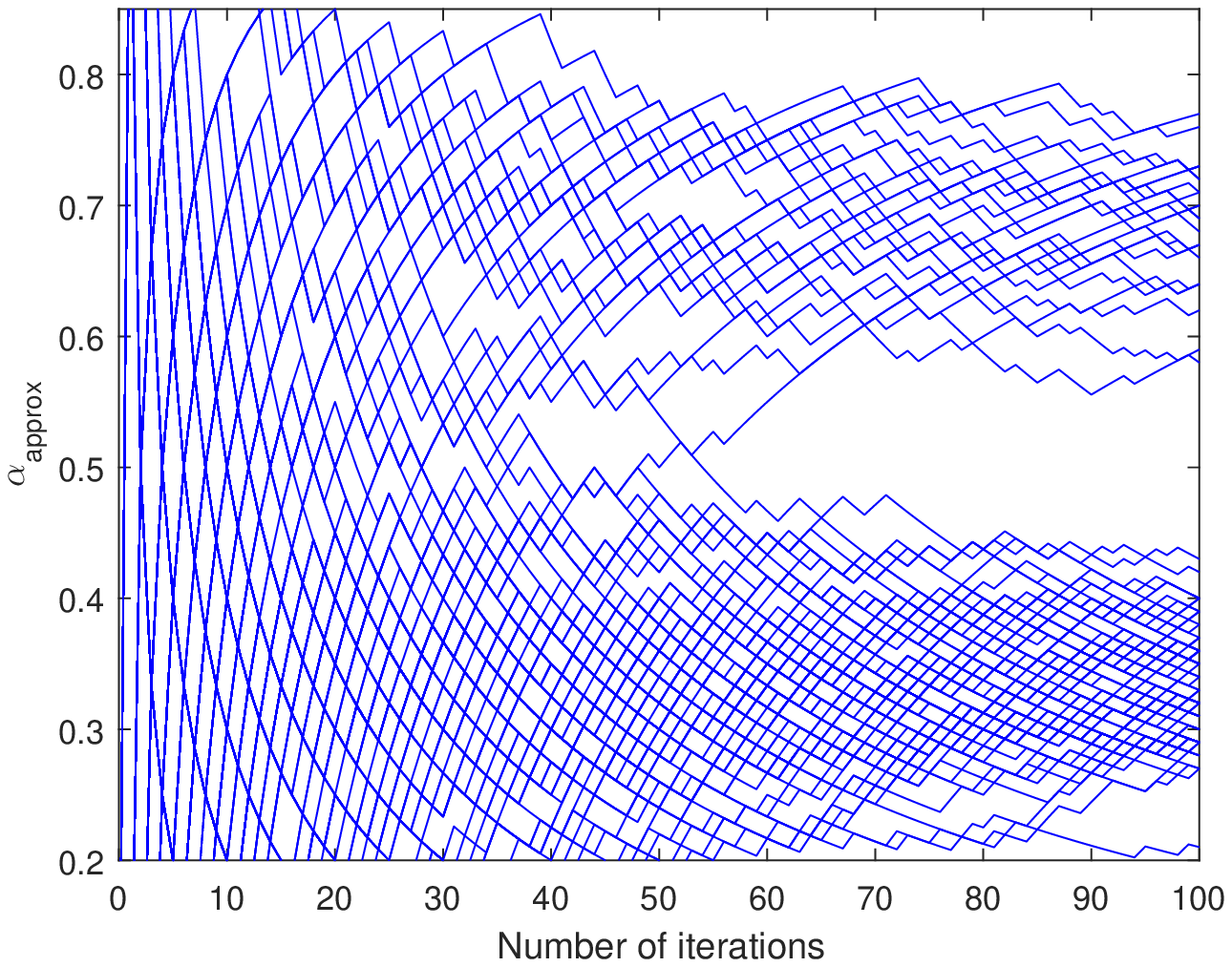} & \includegraphics[scale=0.4]{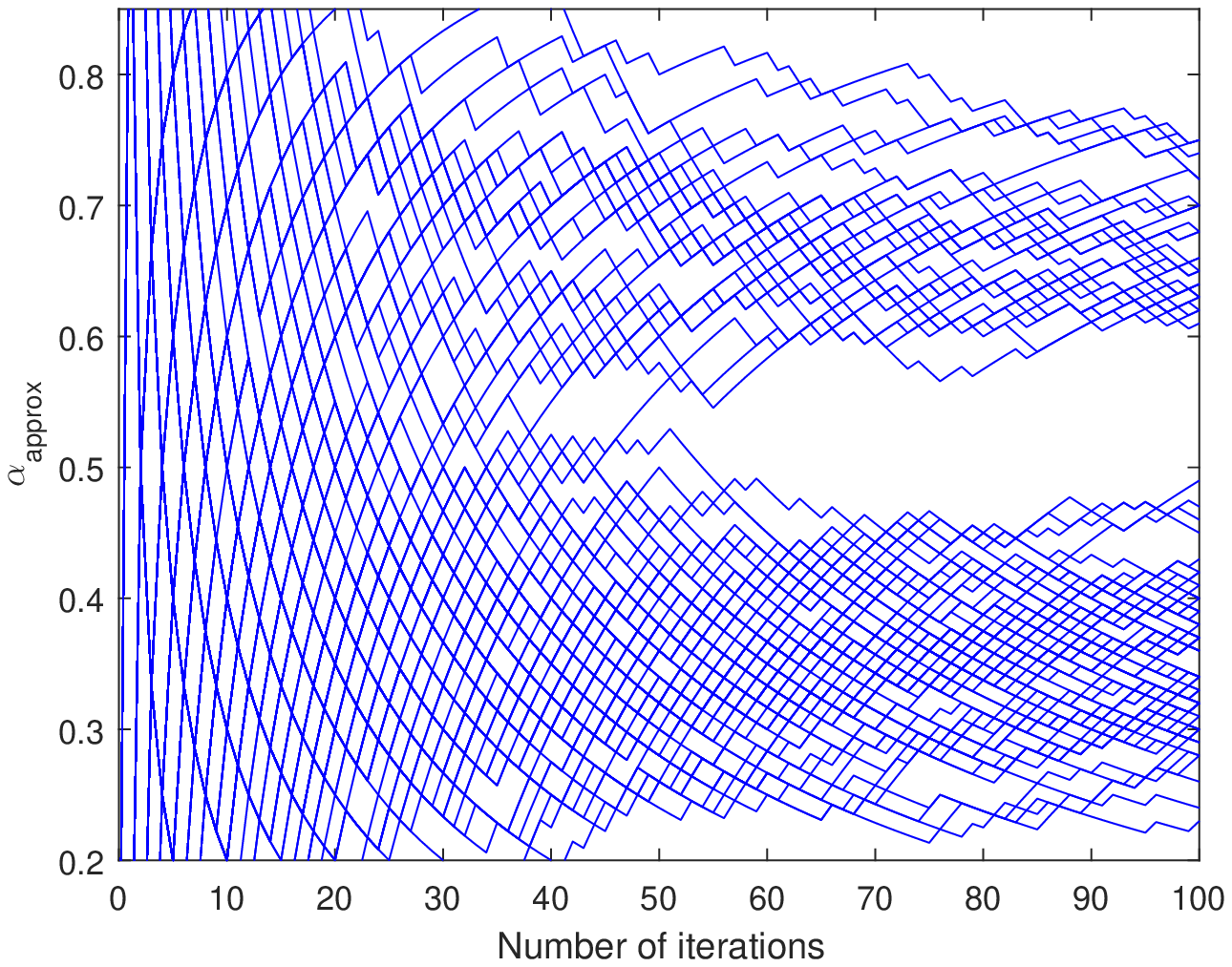}
          	\end{tabular}
          	\captionof{figure}{100 random walks for the cases $\ket{\psi}=\ket{0}$, $\ket{\psi}=\ket{+}$, $\ket{\psi}=\ket{1}$ and $\ket{\psi}=\ket{-}$.}
          \end{center}
      Given classical data 0 or 1, and encoded randomly either in the computational or the Hadamard basis. The impact of the results on the QKD, shows that the security of the system enables an Eavesdropper to determine the sent message with a $\simeq 75\%$ correctness rather than $50\%$ \cite{gisin2002quantum} which is considered as a random guess.
     \section{Conclusion}
     In this paper a quantum algorithm has been proposed to solve the quantum state discrimination problem using partial negation and weak measurement. The proposed approach used the partial negation operator to cause the effect of the weak measurement on an unknown qubit $\ket{\psi}$ which is promised to be in one of the four states $\ket{0}$, $\ket{1}$ (computational basis), $\ket{+}$ or $\ket{-}$ (Hadamard basis), such that the superposition of the qubit will not be destroyed. The algorithm determined the state of the unknown qubit $\ket{\psi}$ approximately by computing an approximated value of its amplitude ${\alpha_{approx}}$ using equation (22), such that after an arbitrary number of iterations it collapses to either 0 or 1. It can be revealed that if ${\alpha_{approx}}$ is biased towards 0, then $\ket{\psi}$ is either in states $\ket{0}$ or $\ket{+}$, otherwise it is in states $\ket{1}$ or $\ket{-}$. The proposed algorithm distinguished between the states $\ket{0}$, $\ket{+}$ and $\ket{1}$, $\ket{-}$ by determining approximately whether the qubit is encoded in the Hadamard or the computational basis, where the value of ${\alpha_{approx}}$ is calculated after 2 iterations, if it lies within a certain interval then, the qubit is decided to be in the Hadamard basis. The algorithm succeeds in determining the state of the unknown qubit approximately with an average of $\simeq 75\%$ for all of the considered states using a single copy. Accordingly, the algorithm can be applied to discover the message on the quantum key distribution, where the Eavesdropper will be able to determine a sent message between two parties through a quantum channel with $\simeq 75\%$ rather than $50\%$ \cite{gisin2002quantum}.
     
	\bibliographystyle{unsrt}
	\bibliography{./manuscript}	
\end{document}